\documentclass[nofootinbib,superscriptaddress,longbibliography,a4paper,twocolumn]{revtex4-1}
\usepackage{graphicx}
\usepackage{dcolumn}
\usepackage{bm,bbm}
\usepackage{xcolor}
\usepackage{tcolorbox}
\usepackage{algorithm}
\usepackage{algpseudocode}
\usepackage{amsmath}
\usepackage{bbold}
\usepackage{braket}
\usepackage{nameref}
\usepackage[toc,page]{appendix}

\usepackage[colorlinks,breaklinks,linkcolor={blue},citecolor={magenta},urlcolor={blue}]{hyperref}

\makeatletter
\newcommand\org@hypertarget{}
\let\org@hypertarget\hypertarget
\renewcommand\hypertarget[2]{%
  \Hy@raisedlink{\org@hypertarget{#1}{}}#2%
  }
\makeatother

\begin{document} 

\title{High-speed imaging of spatiotemporal correlations in Hong-Ou-Mandel interference}

\author{Xiaoqin Gao}
\affiliation{Department of physics, University of Ottawa, Advanced Research Complex, 25 Templeton Street, K1N 6N5, Ottawa, ON, Canada}

\author{Yingwen Zhang}
\email{Yingwen.Zhang@nrc-cnrc.gc.ca}
\affiliation{National Research Council of Canada, 100 Sussex Drive, K1A 0R6, Ottawa, ON, Canada}

\author{Alessio  D'Errico}
\affiliation{Department of physics, University of Ottawa, Advanced Research Complex, 25 Templeton Street, K1N 6N5, Ottawa, ON, Canada}

\author{Khabat Heshami}
\affiliation{National Research Council of Canada, 100 Sussex Drive, K1A 0R6, Ottawa, ON, Canada}
\affiliation{Department of physics, University of Ottawa, Advanced Research Complex, 25 Templeton Street, K1N 6N5, Ottawa, ON, Canada}

\author{Ebrahim Karimi}
\affiliation{Department of physics, University of Ottawa, Advanced Research Complex, 25 Templeton Street, K1N 6N5, Ottawa, ON, Canada}
\affiliation{National Research Council of Canada, 100 Sussex Drive, K1A 0R6, Ottawa, ON, Canada}

\begin{abstract}
The Hong-Ou-Mandel interference effect lies at the heart of many emerging quantum technologies whose performance can be significantly enhanced with increasing numbers of entangled modes one could measure and thus utilize. Photon pairs generated through the process of spontaneous parametric down conversion are known to be entangled in a vast number of modes in the various degrees of freedom (DOF) the photons possess such as time, energy, and momentum, etc. Due to limitations in detection technology and techniques, often only one such DOFs can be effectively measured at a time, resulting in much lost potential. Here, we experimentally demonstrate, with the aid of a time tagging camera, high speed measurement and characterization of two-photon interference. {With a data acquisition time of only a few seconds, we observe a bi-photon interference and coalescence visibility of $\sim64\%$ with potentially up to $\sim2\times10^3$ spatial modes}. These results open up a route for practical applications of using the high dimensionality of spatiotemporal DOF in two-photon interference, and in particular, for quantum sensing and communication.
\end{abstract}

\date{\today}
\maketitle
\section*{Introduction} 
Whenever two identical photons enter a 50:50 beamsplitter (BS) through different input ports, they will ``bunch" together -- a result of bosons wanting to occupy the same quantum state -- and will always exit through the same output port~\cite{PhysRevLett.59.2044, Fearn1989TheoryOT,Rarity:88,Bouchard_2021}. This is known as the two-photon interference effect and was experimentally demonstrated by the Hong-Ou-Mandel (HOM) interference experiment~\cite{PhysRevLett.59.2044}. Two-photon interference arises from the destructive interference between Feynmann paths; hence, it is purely quantum in nature~\cite{pittman1996can}. HOM interferometry is nowadays widely used in quantum optics experiments and related technologies, e.g., for the characterization of photon-pair sources~\cite{graffitti2018design}, in quantum imaging~\cite{nasr2003demonstration, teich2012variations, tang2017spiral, ibarra2020experimental}, quantum communication protocols and quantum cloning \cite{nagali2009optimal,bouchard2017high}, generation of entangled states \cite{kim2003experimental, erhard2018experimental}, for Boson sampling~\cite{tillmann2013experimental}, and the generation of multi-mode N00N state with $N=2$ for quantum supersensitivity~\cite{PhysRevLett.112.103604} and quantum superresolution~\cite{PhysRevLett.112.223602}. The figure of merit of HOM interference is the visibility of the photon ``bunching'' effect, either observed as a dip in the two-photon coincidence between photons exiting from different ports of the BS, or as a peak in the two-photon coincidence between photons exiting from the same ports of the BS, referred to as coalescence. The visibility of the HOM peak/dip for a classical source cannot be higher than 50\%~\cite{Kaltenbaek2006,Bouchard_2021}. The visibility of HOM experiments is affected by all the imperfections which introduce differences between the quantum state of the two incoming photons, e.g., differences in frequency, polarization, and transverse momentum. Consequently, experiments typically involve interference between two photons which are superpositions of a small number of spatial and temporal modes. This is done either by post-selection, using single-mode optical fibres in the detection stage, or by filtering the spatial modes with pinholes~\cite{Walborn_2003}. However, there is a strong interest in quantum optics applications involving a large number of modes: high-dimensional states exhibit large information capacity \cite{Erhard_2020}, provide better security thresholds in quantum communication protocols~\cite{cerf2002security}, and can be useful for reducing the noise in quantum imaging~\cite{brida2011experimental}. 
{Two-photon interference has been previously investigated with spatial modes, e.g., Hermite-Gauss modes \cite{Walborn_2003, PhysRevA.94.033855}, Laguerre-Gauss modes \cite{PhysRevLett.104.020505,PhysRevA.89.013829, Zhange1501165}, and position and momentum modes \cite{PhysRevA.73.063827, PhysRevX.10.031031}. } 

\begin{figure*}[t]
\includegraphics [width= 1\linewidth]{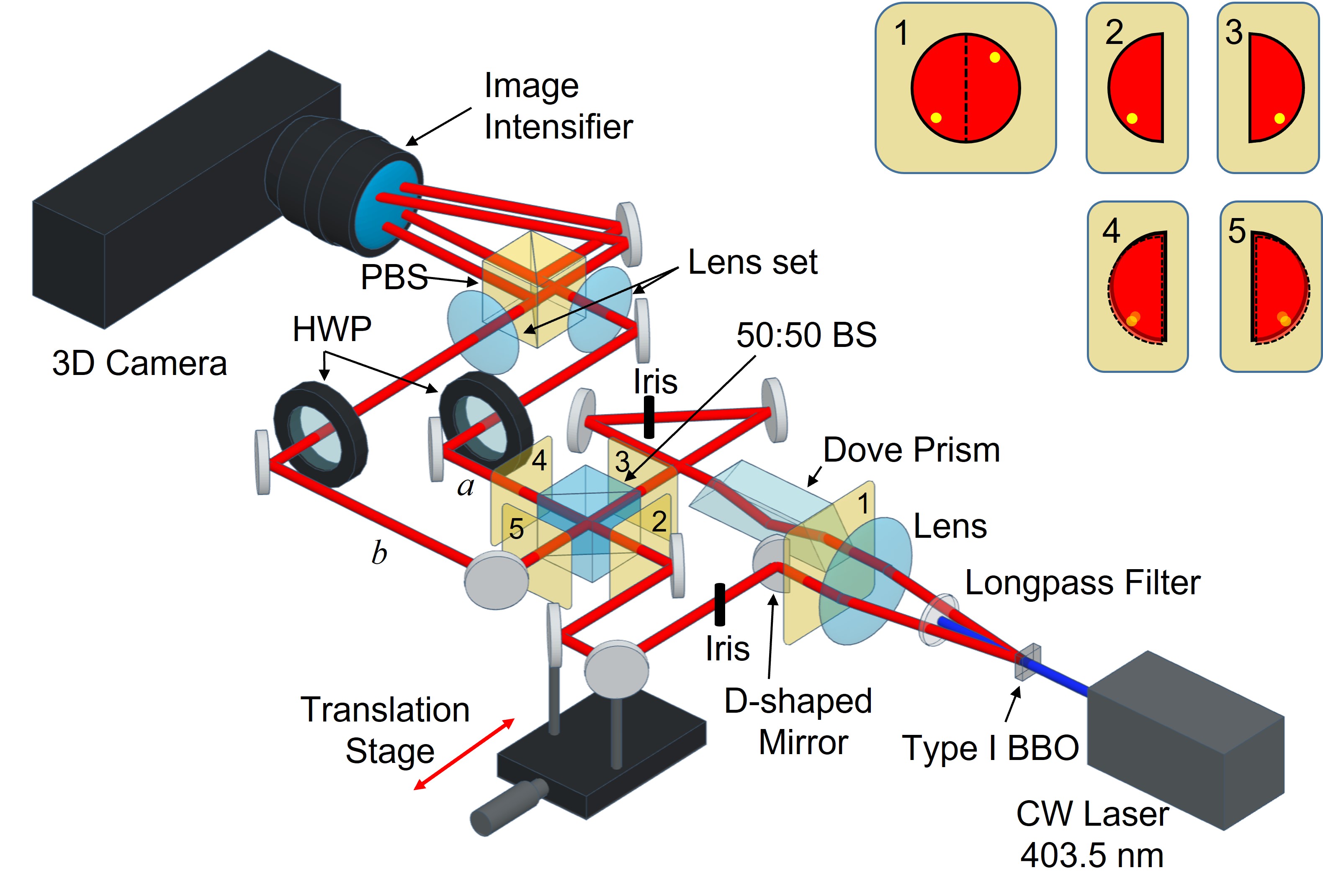}
\centering
\caption{Experimental setup for imaging spatiotemporal HOM interference in the near field of the BBO crystal. BS-beamsplitter with outputs labelled ``$a$'' and ``$b$", PBS-polarizing beamsplitter, HWP-half-wave plates oriented at $22.5^{\circ}$. The insets show the beam projections at (1) before the D-shaped mirror, {(2) and (3) before the BS, and (4) and (5) after the BS, respectively.}}
\label{fig:setup}
\vspace{0.2cm}%
\end{figure*}

Performing HOM measurements in the spatiotemporal domain requires very long data acquisition times, which often have limited accuracy. Either physical raster scanning of detectors~\cite{Lubin:19} and apertures~\cite{PhysRevA.73.063827} are employed, or spatially resolved EMCCD cameras are used \cite{PhysRevX.10.031031}. The former technique generally provides good timing but poor spatial resolution, while the latter gives good spatial resolution but poor timing and significant noise. In this work, we make use of a time-tagging camera technology (TPX3CAM \cite{3D,Nomerotski2019}) to image HOM interference in the spatiotemporal domain. Having a total of $256\times256$ pixels with a pixel size of 55\,$\mu$m and an effective per pixel timing resolution of $\sim 6$\,ns, allows the user to surpass the trade-off between high spatial or high timing resolution in photon correlation measurements. Spatiotemporal correlation measurements are performed in the near field of the crystal with a peak and dip visibility of, respectively, $\sim68\%$ and $\sim 62\%$. {By reducing the region of interest (ROI) down to roughly 1/3 the beam waist through post-processing we can achieve the dip and peak visibility to $\sim 86\%$ and $\sim 94\%$, respectively. We also show that up to $\sim2000$ spatial modes can potentially be measured for a well designed Spontaneous Parametric Down Conversion (SPDC) source with high position/momentum correlation.} The same camera system have also been recently used to observe HOM interference in the spectral-temporal domain~\cite{Zhang2021}.

\section*{Results}
In order to simultaneously perform HOM interference between a high number of modes, we consider the quantum state generated in Type-I SPDC. The bi-photon quantum state, in the thin crystal approximation, is given by,
\begin{equation}
    \ket{\Psi}=\int d^2\mathbf{q}\, \phi(\mathbf{q})\,\ket{\mathbf{q}}_s\ket{\mathbf{-q}}_i,
    \label{SPDCstate}
\end{equation}
where $\mathbf{q}=(q_x,q_y)$ is the projection of the linear momentum on the plane transverse to the propagation direction, $\phi(\mathbf{q})$ is the bi-photon wavefunction (determined by the pump amplitude shape and the phase matching) \cite{walborn2010spatial}, and $i$ and $s$ stand for idler and signal photon, respectively (in this expression, we have neglected the very low possibility for the generation of multiple photon pairs with the same spatiotemporal modes).  Equation~\eqref{SPDCstate} can be seen as a consequence of momentum conservation: any signal photon emitted with momentum $\mathbf{q}$ is correlated with an idler photon possessing momentum $-\mathbf{q}$. Transverse momentum is mapped into transverse position, and \emph{vice versa}, using a lens where the nonlinear crystal is kept at the lens focal plane.

As illustrated in Fig.~\ref{fig:setup}, correlated photon pairs, generated through SPDC by pumping a 0.5-mm-thick Type-I BBO crystal with a 403.5\,nm continuous-wave laser, are routed into two different paths by a D-shaped mirror. In order to obtain HOM interference for each value of $\mathbf{q}$, the two paths of the interferometer must be designed in such a way that, at the output ports of the BS, one of the two paths has implemented the coordinates transformation $\mathbf{q}\rightarrow -\mathbf{q}$. This is obtained by means of a vertically oriented Dove prism, implementing $q_y\rightarrow -q_y$, and by a different parity in the number of reflections in the two paths (including the reflection from the BS), which corresponds to applying $q_x\rightarrow -q_x$. An alternative approach would be to exploit the transverse position correlations, which would require imaging the crystal plane on the BS (e.g., see~\cite{PhysRevX.10.031031}); however, the resulting visibility can be strongly affected by imaging imperfections.  A detailed experimental setup is shown in Fig.~\ref{fig:setup}. Two-photon interference is obtained at the outputs of a 50:50 BS, with the outputs labelled ``$a$'' and ``$b$". Two half-wave plates (HWPs) oriented at $22.5^{\circ}$, and a polarizing beamsplitter are employed to further split the two beams emerging from the 50:50 BS into four, and hence allowing the observation of the HOM dip and peak simultaneously. Using different lens sets we can image the SPDC photons in either the near field or far field of the BBO crystal. 

When perfect multi-mode HOM interference is registered, the resulting state (neglecting the contributions coming from emission of 4 or more photons), in the photon number basis, is a multi-mode two-photon N00N state, i.e.,
\begin{equation}
    \ket{\psi}=\mathcal{N}\,\sum_{k=1}^M \left(\ket{ 2_{k,a}, 0_{k,b}}+\ket{ 0_{k,a}, 2_{k,b}}\right)\otimes\ket{\text{vacuum}'}.
    \label{eq:N00N}
\end{equation}
Here, $\mathcal{N}$ is a normalization factor, $k$ is the mode index, $M$ is the total number of modes, $a/b$ are the BS output ports indices, and $\ket{\text{vacuum}'}$ represents vacuum state for all other modes $k'\neq k$. The choice of the spatial mode set depends on the actual application that one may consider. In imaging experiments, $k$ can be chosen as corresponding to the camera's pixel position. Therefore, it depends on the imaging system and can be associated with an area of the transverse position or transverse momentum space. This choice is not arbitrary since one can always devise a protocol in which pixel degrees of freedom are consistently mapped into spatial ones. 

\begin{figure*}[tbph]
\includegraphics [width= 1\textwidth]{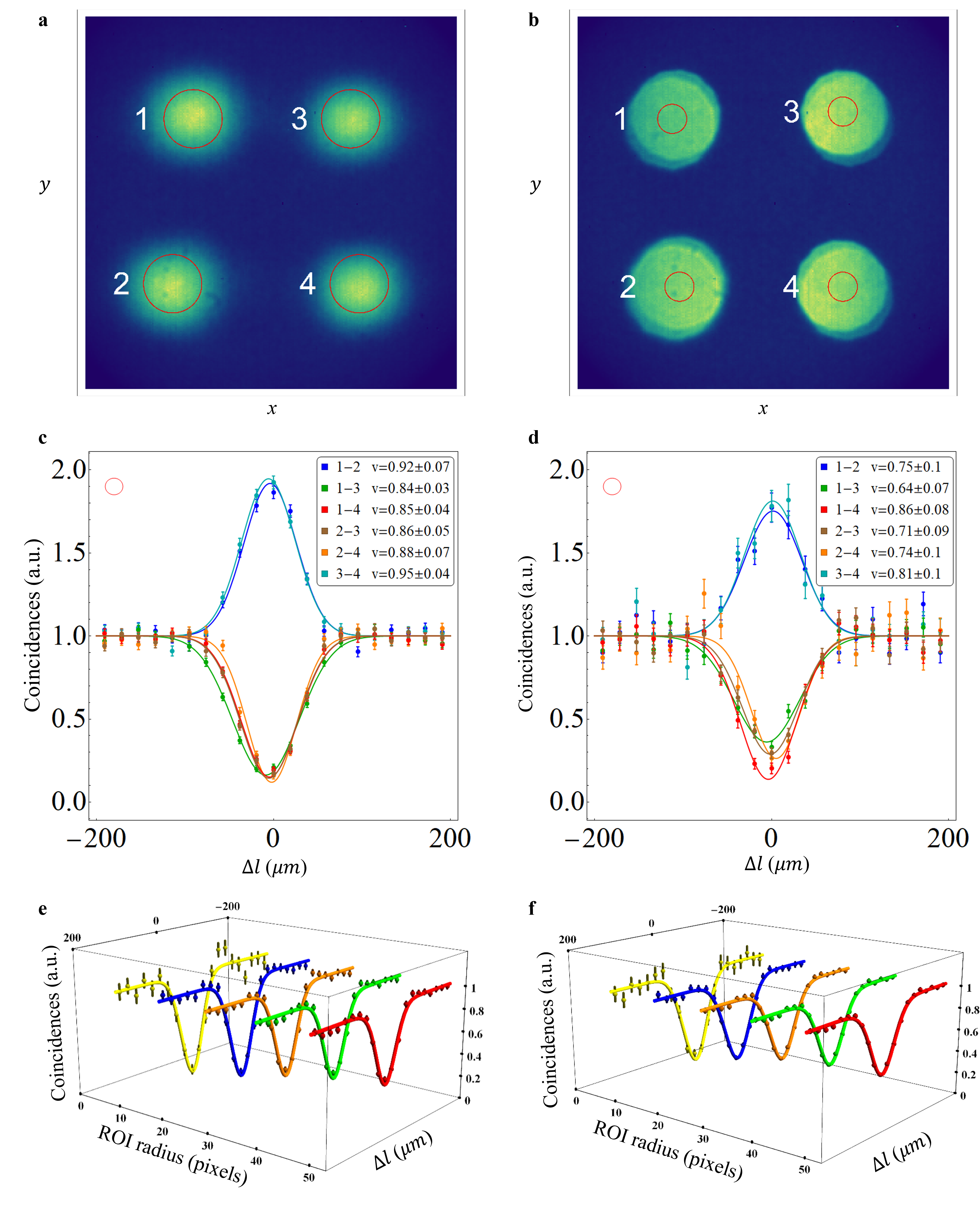}
\centering
\caption{Imaging of HOM interference in both the near field and the far field of the BBO crystal. \textbf{a(b)} Continuous exposure images recorded on the camera placed in the near (far) field. \textbf{c(d)} Two-fold normalized coincidences in the near (far) field {as a function of ${\Delta}l$, the path length difference between the two photons before the BS, with selecting a circular ROI with a radius of 20(10) pixels}. The different colored data points and Gaussian fits corresponds to coincidences between different spots (specified in the figure legend). Cases 1-2 and 3-4 correspond to photon pairs exiting the same port of the interferometer. All other combinations correspond to photon pairs exiting different ports of the BS. \textbf{e(f)} Two-fold normalized coincidences in the near (far) field when selecting ROI of different radius. {With a ROI radius of 10, 20, 30, 40 and 50 pixels selected between the beam spot 2-4 (1-4), the corresponding visibilities of the HOM dip are $0.92 \pm 0.10$ ($0.86 \pm 0.08$), $0.88 \pm 0.07$ ($0.75 \pm 0.04$), $0.78 \pm 0.06$ ($0.65 \pm 0.03$), $0.70 \pm 0.05$ ($0.61 \pm 0.02$), and $0.64 \pm 0.05$ ($0.60 \pm 0.02$) respectively. The beam in \textbf{a} appears circular because it is imaged from the image plane of the crystal where the beam is circular. \textbf{b} appears circular because two iris with apertures of 4mm were placed after the D-shaped mirror, which selects out a circular shaped region of the D-shaped beam.} All analysis were performed from a single data set taken over 10\,s.
}
\label{fig:nearfourbeam4mm}
\end{figure*}
\begin{figure}[tbph!]
\includegraphics [width=0.8\linewidth]{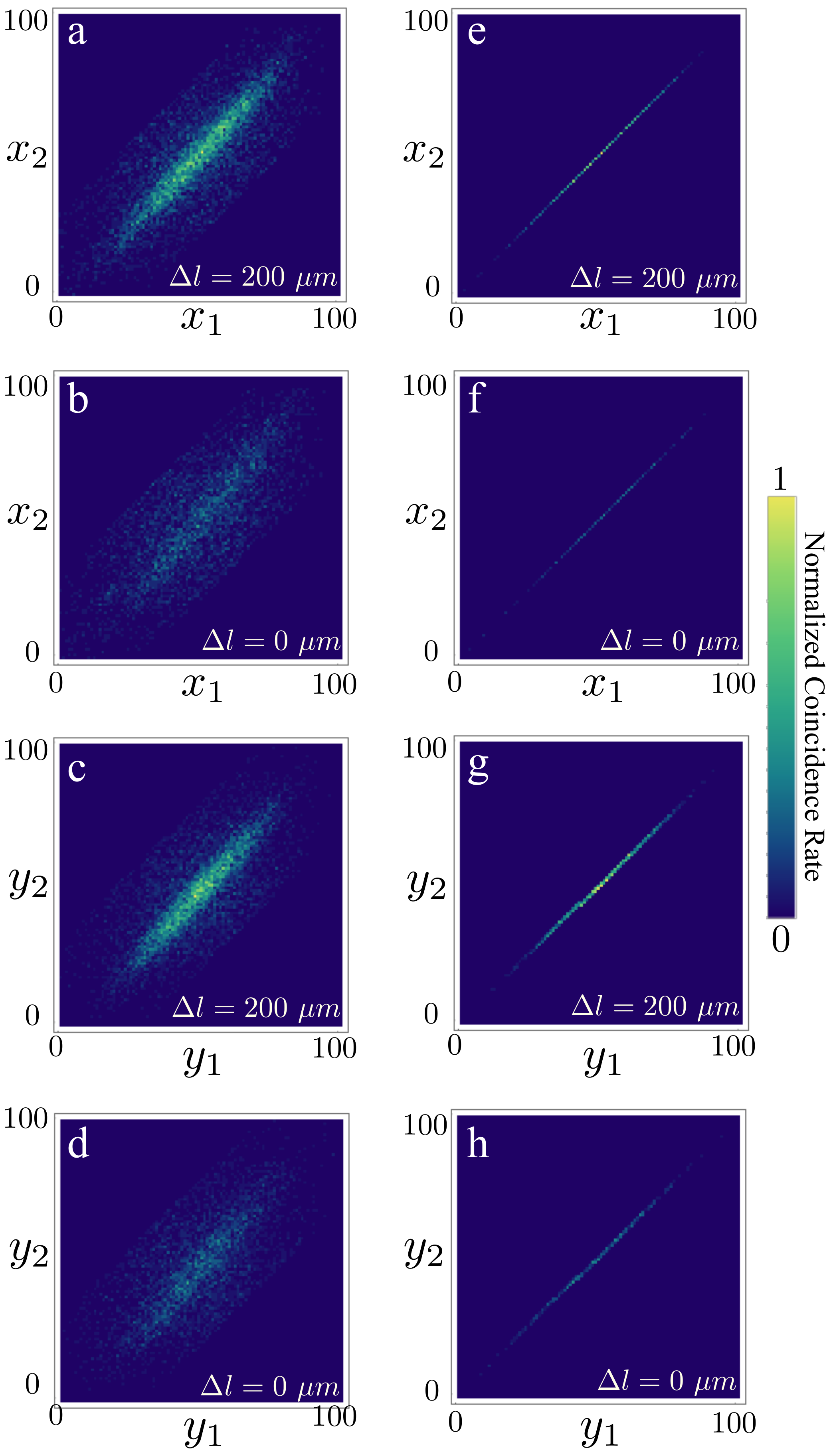}
\centering
\caption{Measured near field spatial correlations. Through post-processing, the time-tagging camera allows one to filter out the counts from a given region in the correlation space. With a 50 pixel radius ROI and a 30 pixel wide spatial correlation selection band (\textbf{a--d}) we obtain a HOM dip visibility of $0.59\pm0.03$. However, with a 1 pixel wide selection band (\textbf{e--f}) we obtain a visibility of $=0.67\pm0.04$.  {${\Delta}l$ represents the path length difference between the two photons before the BS.}}
\label{fig:fartwobeam4mm}
\end{figure}

The results from imaging HOM interference, taken over a period of 10\,s, in both the near field and the far field of the BBO crystal are shown in Fig.~\ref{fig:nearfourbeam4mm}.  Fig.~\ref{fig:nearfourbeam4mm}\,\textbf{a} (\textbf{b}) shows the images of the beam spots in the near (far) field of the crystal captured on camera through continuous exposure with the beam spots 1--2 and 3--4 coming from, respectively, the BS output ports $a$ and $b$. Hence, coincidences between either spots 1 and 2, or 3 and 4 will show coalescence peaks while all the other combinations give HOM dips as seen in Fig.~\ref{fig:nearfourbeam4mm}\,\textbf{c} (\textbf{d}). We measured total coincidences within a circular region of interest (ROI) around each spot with a 20 pixel radius and obtained an average visibility of $\sim88\%$ ($\sim 75\%$). The visibility reduces with a larger ROI as seen in Fig.~\ref{fig:nearfourbeam4mm}\,\textbf{e} (\textbf{f}), this is likely a result of misalignment in the beam overlap becoming more discernible at the outer regions of the beams. We see that a  visibility of $\sim64\%$ ($\sim60\%$) can be achieved for a HOM dip between spot 2-4 when the ROI radius is equal to 50 pixels.  {Whereas with a smaller ROI radius of 20 pixels, we obtain a visibility of $\sim88\%$ ($\sim75\%$).} Thus a larger ROI would include a larger number of modes affected by the setup imperfections. A similar analysis on the influence of the selected region size on the visibility for a HOM peak can be found in the supplementary.

As a next step, we examined how the HOM interference is affected by the spatial correlations. Figure~\ref{fig:fartwobeam4mm}\,\textbf{{a}}-\textbf{{d}} show the joint probability distribution (JPD) of the spatial correlations in both the horizontal and vertical directions. We measured the joint probability distributions in two different scenarios: outside the HOM dip (Fig.~\ref{fig:fartwobeam4mm}\,\textbf{a} and \textbf{{c}}), and inside the HOM dip (Fig.~\ref{fig:fartwobeam4mm}\,\textbf{{b}} and \textbf{{d}}). The state generated in the latter scenario is a 2 photon N00N state.  In post-selection, we are free to choose the width of this spatial correlation band, i.e. how far away the spatial coordinates of a coincidence event must be from the central diagonal of the JPD to be considered spatially correlated. By placing a stricter condition (at the cost of losing coincidence events), such as considering only events on the diagonal to be spatially correlated, as shown inFig.~\ref{fig:fartwobeam4mm}\,\textbf{{e}}-\textbf{{h}}, we can reduce the background noise and improves the HOM dip visibility by $\sim 10\%$. The visibility of bi-photon interference, and consequently the N00N state generation fidelity, depends on the indistinguishability of the photons at the BS.  {Another thing to note here is that even slight rotations between the two beams will cause a spatial mismatch and reduce the HOM visibility. The effect of this becomes worse as one moves towards the edge of the beam. By placing an iris with a variable aperture after the BBO crystal will block off the outer edges of the far field beam and thus reducing the effect of the spatial mismatch from slight beam rotations and thus further improving HOM visibility.} We have also performed the same analysis with the camera placed in the crystal's far-field. The obtained results are similar to the near-field and are shown in the supplementary.

\begin{figure}[tbph!]
\includegraphics [width=0.8\linewidth]{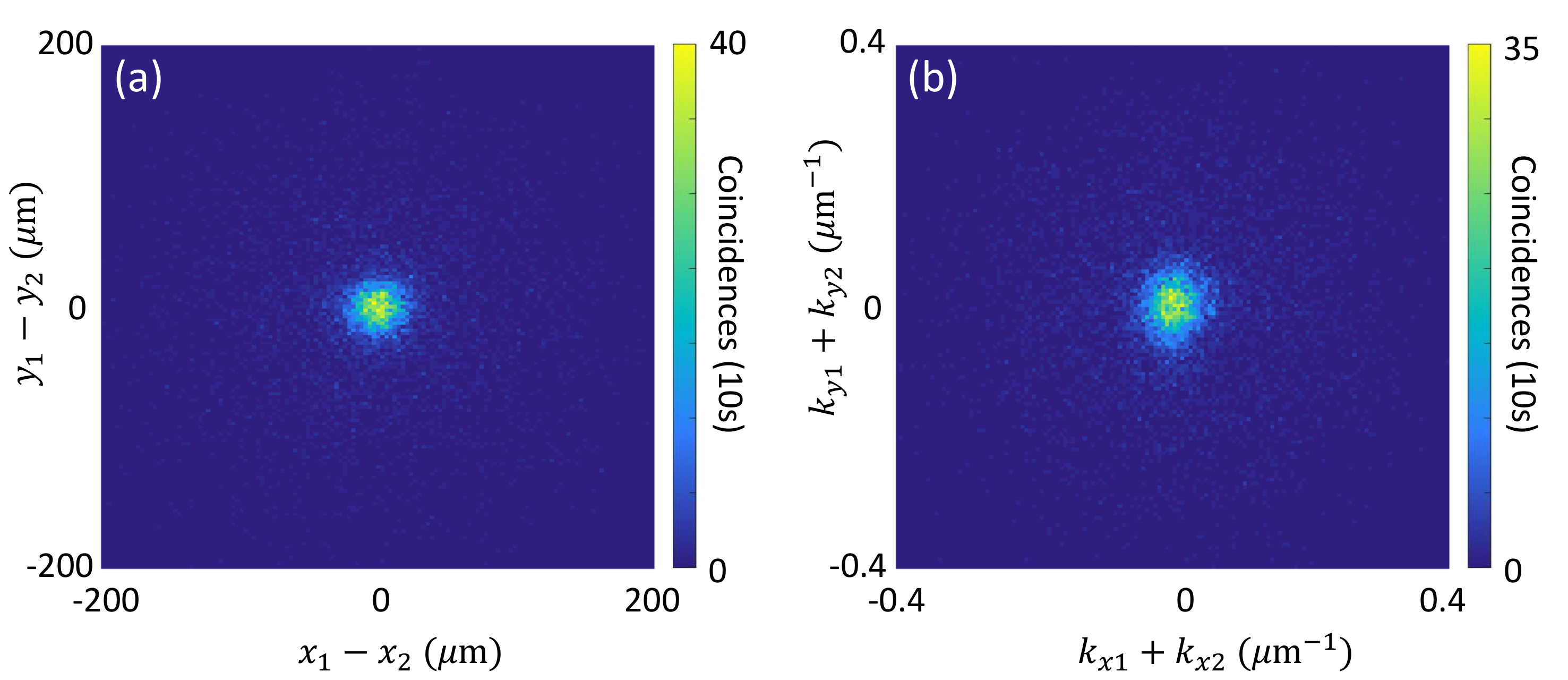}
\centering
\caption{(a) The difference coordinate projection showing the measured position correlation of the SPDC source. (b) The sum coordinate projection showing the measured momentum correlation of the SPDC source.   }
\label{fig:fig4}
\end{figure}

{Given the beam shape of SPDC seen in Fig.~\ref{fig:nearfourbeam4mm} and the degree of position/momentum correlation, the effective number of spatial modes is not necessarily the same as the number of pixels within the ROI. We therefore define the effective number of spatial modes as
\begin{equation}
    N_s =  \frac{\sum_iI_i}{4\sigma^2 I_\text{max}},
\label{Schmidt}
\end{equation}
where $I_i$ is the number of photons detected on pixel $i$ of the beam seen in Fig.~\ref{fig:nearfourbeam4mm}\,\textbf{a}(\textbf{b}), $I_\text{max}$ is the maximum number of photons detected on a pixel in the beam and $\sigma$ is the position correlation width in pixels (or momentum correlation if imaging the momentum plane) given by a Gaussian fit of $f(x,y) = a \exp\left(-\frac{(x-b_1)^2+(y-b_2)^2}{2\sigma}\right)+d$ to the difference and sum coordinate projection plots for the position and momentum degrees of freedom as shown in Fig.~\ref{fig:fig4}. From fitting the Gaussian function to Fig.~\ref{fig:fig4}, we obtained a $\sigma = 3.73(4.12)$ pixels for the position(momentum) correlation which gives us a total of $31(38)$ effective spatial modes for a 50 pixel radius ROI in the near(far) field and $15(14)$ effective spatial modes for a 20 pixel radius ROI. With a well designed SPDC source with a position(momentum) correlation of 1 pixel wide the number of effective spatial modes can be boosted up to $\sim1.7(2.6)\times10^3$ for a 50 pixel radius ROI and $\sim8.6(9.8)\times10^2$ for a 20 pixel radius ROI.} 

{We can also estimate the effective number of entangled spatial modes, the Schmidt number, generated by the SPDC source through measuring the degree of correlation in the position and momentum DOFs. Theoretically this can be determined through the physical features of the pump laser and SPDC crystal geometry. By using a Gaussian pump and a Gaussian approximation for the sinc shaped phase matching function~\cite{Law_2004,Defienne2019}, the biphoton amplitude is given by 
\begin{equation}
    \Psi(\textbf{k}_1,\textbf{k}_2) \propto \exp\left(-\frac{\sigma_r^2|\textbf{k}_1-\textbf{k}_2|^2}{2}\right)\exp\left(-\frac{|\textbf{k}_1+\textbf{k}_2|^2}{2\sigma_k^2}\right).
\end{equation}
}

{The position-correlation width $\sigma_r$ depends on the crystal length $L$ and the pump wavelength $\lambda_p$ as given by $\sigma_r=\sqrt{\alpha L\lambda_p/(2\pi)}$, with $\alpha = 0.455$ a scaling constant for approximating the sinc function with a Gaussian~\cite{Chan2007}. $\sigma_k$ is the momentum-correlation width given by $\sigma_k = \sqrt{1/l_c^2+1/(4\omega_p^2)}$, with $l_c$ the coherence length of the pump laser and $\omega_p$ the pump beam waist radius. The Schmidt number $K$ can be given in terms of $\sigma_r$ and $\sigma_k$~\cite{Law_2004} as
\begin{equation}
    K = \frac{1}{4}\left(\frac{1}{\sigma_r\sigma_k} + \sigma_r\sigma_k\right)^2.
\end{equation}
}

{Using our experimental parameters of $L=0.5$\,mm; $\lambda_p=403.5$\,nm; $l_c\approx50$\,$\mu$m; $\omega_p\approx75$\,$\mu$m, we obtained a $\sigma_r = 3.8$\,$\mu$m and $\sigma_k = 0.021$\,$\mu$m$^{-1}$, thus giving us a $K=39$. 
}

{Experimentally, by fitting a Gaussian function of the form $f(x,y) = a \exp\left(-\frac{(x-b_1)^2+(y-b_2)^2}{2\sigma_{r,k}^2}\right)+d$ to the difference and sum coordinate projection plots for the position and momentum degrees of freedom as seen in Fig.~\ref{fig:fig4}, we obtained a $\sigma_r=10$\,$\mu$m and $\sigma_k=0.023$\,$\mu$m$^{-1}$ thus giving a $K=4.8$. The discrepancy between the measured and expected $K$ lies in that the camera was not placed exactly on the image plane of the crystal thus resulting in a measured $\sigma_r$ much larger than expected. However, we must note that by measuring the degree of correlation will only give us an estimate of the Schmidt number of the source, to know what is the actual number of Schmidt modes measurable by the measurement system, one will have to measure individually each Schmidt mode in either the Laguerre or Hermite Gaussian basis.}

\section*{Conclusion and Outlook}
{In the previous approaches of two-photon interference with spatial modes \cite{Walborn_2003, PhysRevA.94.033855, PhysRevLett.104.020505,PhysRevA.89.013829, Zhange1501165, PhysRevA.73.063827, PhysRevX.10.031031}}, each spatial mode is scanned individually, or the spatial modes are measured simultaneously but with a poor timing resolution. Moreover, these techniques require very long data acquisition times. Here, we have demonstrated simultaneous measurements of spatial and temporal correlations in HOM interference with high resolution and high speed using a time-tagging camera. 

Our ability to swiftly measure a large number of spatiotemporal modes in HOM interference can be exploited in many applications, such as in high-dimensional device independent quantum key distribution where the position and momentum DOF can be used as the two mutually unbiased basis, a source for high-dimensional two-photon N00N states, to be used in sensing applications~\cite{hiekkamaki2021photonic}, and quantum imaging applications such as in optical coherence tomography \cite{teich2012variations}.

\textbf{Note.} During the submission of this manuscript we have recently become aware of the work in Ref. \cite{defienne2021hong}, where a similar setup as the one we implemented has been used for depth imaging.

\section*{Funding}
This work was supported by the Canada Research Chairs (CRC), the High Throughput and Secure Networks (HTSN) Challenge Program at the National Research Council of Canada, Canada First Research Excellence Fund (CFREF) Program, and Joint Centre for Extreme Photonics (JCEP). 

\section*{Acknowledgments}
The authors would like to thank Aephraim Steinberg, Fr\'ed\'eric Bouchard, Dilip Paneru and Alicia Sit for valuable discussions, and Manuel F Ferrer-Garcia for the great help with figures, and Mohammadreza Rezaee for the great help in the laboratory management.




\bibliography{refs}

\newpage
\clearpage
\onecolumngrid

\section*{Supplementary\\High-speed imaging of spatiotemporal correlations in Hong-Ou-Mandel interference}

\begin{figure}[tbph]
\includegraphics [width= 1\textwidth]{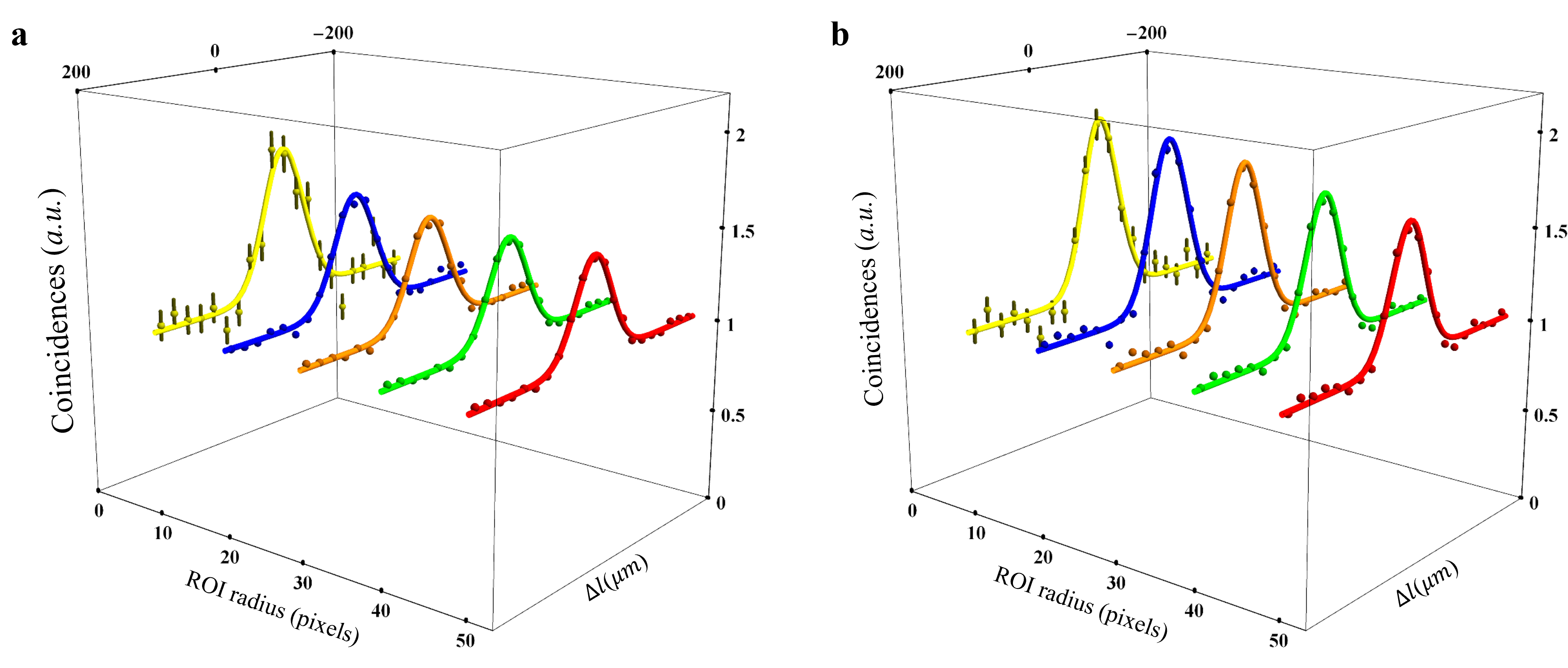}
\centering
\vspace{-0.3cm}%
\caption{
{\bf Visibilities for the HOM peak in the far field and the near field with a ROI radius of 10, 20, 30, 40 and 50 pixels respectively.}
\textbf{a} Multi-mode HOM interference between the spot 3-4 in the far field. The related visibilities for HOM peak are  $0.81 \pm 0.10$, $0.62 \pm 0.04$, $0.57 \pm 0.04$, $0.54 \pm 0.02$, and $0.53 \pm 0.02$ respectively.
\textbf{b} Multi-mode HOM interference between the spot 1-2 in the near field. 
The related visibilities for HOM peak are  $0.97 \pm 0.10$, $0.92 \pm 0.07$, $0.85 \pm 0.05$, $0.76 \pm 0.05$, and $0.69 \pm 0.05$ respectively.
Some error bars are not visible, being smaller than the symbols.
${\Delta}l$ represents the path length difference between the two photons before the BS.}
\label{fig:farsize}
\vspace{0.2cm}%
\end{figure}

Fig.~\ref{fig:farsize} shows the influence of the selected region radius on the visibility for a HOM peak in both the far field (Fig.~\ref{fig:farsize} \textbf{a}) and the near field (Fig.~\ref{fig:farsize} \textbf{b}).

\begin{figure*}[tbph]
\includegraphics [width= 1\textwidth]{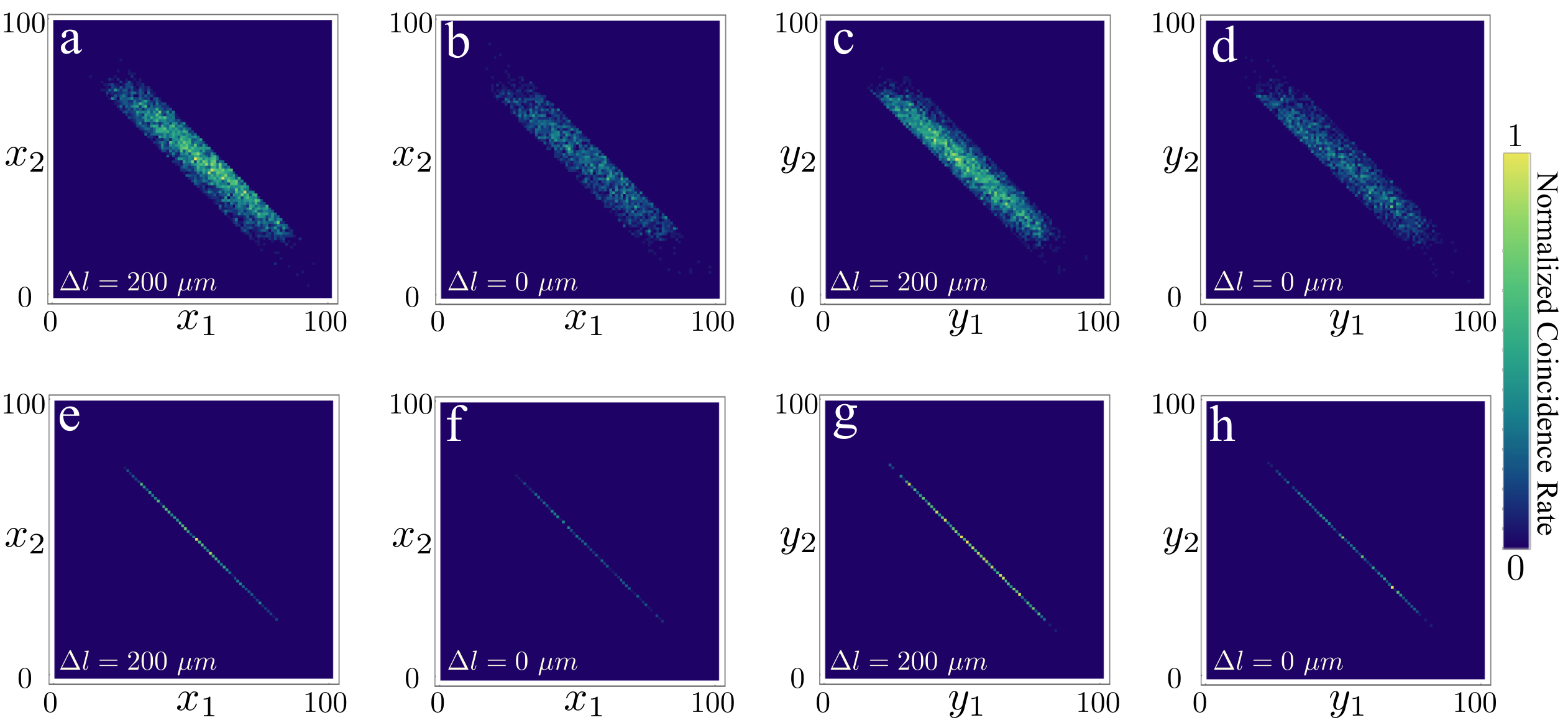}
\centering
\vspace{-0.3cm}%
\caption{
{\bf Detected spatial correlations in the far field.} 
One can reduce background by selecting only diagonal correlations (panels \textbf{{e}}-\textbf{{h}}, corresponding to a 1 pixel wide window). 
The correlations in the $x$ and $y$ coordinate components imaged outside and inside the HOM dip position (corresponding to $\Delta l=0\, \mu$m). 
We have for the 10 pixels (panels \textbf{{a}}-\textbf{{d}}) and 1 pixel selection band between the spots 2–3 (panels \textbf{{e}}-\textbf{{h}}), respectively, ${\cal V}=0.51\pm0.03$ and ${\cal V}=0.56\pm0.06$.
${\Delta}l$ represents the path length difference between the two photons before the BS.}
\label{fig:far}
\vspace{0.2cm}%
\end{figure*}

Thanks to its ability to measure spatiotemporally resolved coincidences, the time-tagging camera allows to directly reconstruct the position correlations of the measured state. As a consequence one can select regions of interest in the correlation space, which allows reducing the background and to post-select on a desired quantum state. In Fig.~\ref{fig:far} we illustrate how one can post-select on the state given in the main text (which manifests perfect momentum anti-correlations, associated with perfect position correlations) within the precision given by the pixel area. Fig. \ref{fig:far} \textbf{{a}}-\textbf{{d}} show the measured correlations (for path delays corresponding to inside and outside the HOM interference) when selecting a circular region of 50 pixels radius and considering counts within a 10 pixels wide stripe in the correlation space. In Fig. \ref{fig:far} \textbf{{e}}-\textbf{{h}} the stripe is reduced to a width of 1 pixel.
In the case shown this amounts to an increase of the visibility of $\sim 10\%$.

\end{document}